# Correlated Noise in the *COBE* [1] DMR Sky Maps

C. H. Lineweaver[2,3], G.F. Smoot[2], C.L. Bennett[4], E.L. Wright[5], L. Tenorio[2],
A. Kogut[6], P.B. Keegstra[6], G. Hinshaw[6], and A.J. Banday[7]

[2] Lawrence Berkeley Laboratory, Space Sciences Laboratory Center for Particle Astrophysics, Building 50-351, University of California, Berkeley CA 94720.

[3] e-mail: lineweaver@astmag.lbl.gov

[4] NASA Goddard Space Flight Center, Code 685, Greenbelt MD 20771.

[5] UCLA Astronomy Department, Los Angeles CA 90024-1562.

[6] Hughes STX Corporation, Code 685.3, NASA/GSFC, Greenbelt MD 20771.

[7] Universities Space Research Association, Code 685.9, NASA/GSFC, Greenbelt MD 20771.

## ABSTRACT

The *COBE* DMR sky maps contain low-level correlated noise. We obtain estimates of the amplitude and pattern of the correlated noise from three techniques: angular averages of the covariance matrix, Monte Carlo simulations of two-point correlation functions, and direct analysis of the DMR maps. The results from the three methods are mutually consistent. The noise covariance matrix of a DMR sky map is diagonal to an accuracy of better than 1%. For a given sky pixel, the dominant noise covariance occurs with the ring of pixels at an angular separation of 60° due to the 60° separation of the DMR horns. The mean covariance at 60° is $0.45\%^{+0.18}_{-0.14}$ of the mean variance. Additionally, the variance in a given pixel is 0.7% greater than would be expected from a single beam experiment with the same noise properties. Auto-correlation functions suffer from a $\sim 1.5\ \sigma$ positive bias at 60° while cross-correlations have no bias. Published *COBE* DMR results are not significantly affected by correlated noise.

*Subject headings:* cosmic microwave background — cosmology: observations

## 1. Introduction

The *COBE* collaboration has reported the detection of anisotropies in the cosmic microwave background radiation (Smoot et al. 1992, Bennett et al. 1994). These anisotropies cannot be attributed to known systematic effects (Kogut et al. 1992) or any known Galactic or extragalactic foreground (Bennett et al. 1992, 1993). Concern about the magnitude of correlated noise in the sky maps has led to this investigation. The DMR measures temperature differences across the sky using pairs of horn antennas with a fixed 60° separation angle (Smoot et al. 1990). The construction of a DMR temperature map from measurements of temperature differences involves a least-squares fit during which essentially uncorrelated measurement errors become correlated temperature errors in the sky map. The structure of this correlated noise is fully described by the pixel temperature covariance matrix which depends only on the 60° horn separation angle and the details of the sky coverage. In this paper we quantify the amplitude and angular dependence of the correlated noise. To first order, the noise in the DMR maps is uncorrelated. The dominant correlation in each pixel is due to the average noise in a ring 60° away. The mean covariance at 60° is 0.45% of the mean variance. This level of correlation will slightly bias auto-correlation functions and their derivatives but previously published DMR results are not significantly affected by correlated noise. The dominant correlated noise term in the DMR maps was discussed briefly in Wright et al. (1994a,b).

The map-making procedure is described in §2. In §3 we use three methods to obtain estimates of the correlated noise in DMR maps and in §4 we discuss the effect of the auto-correlation bias on determinations of the cosmological parameters $n$, $Q_{rms-PS}$ and the rms temperature fluctuation amplitude.

## 2. Map-Making Procedure

The sky is divided into 6144 approximately equal-area pixels. During the $m$th observation ($1 \leq m \leq m_{tot}$ where $m_{tot}$ is the the total number of observations for the channel under consideration), the positive horn is pointing at some pixel $i$, while the negative horn is pointing 60° away at a pixel $j$. The $m$th observation can be modeled as $D_m = \Delta T_m + \Delta_m$, where $\Delta T_m$ is the true temperature difference $T_{i(m)} - T_{j(m)}$, and $\Delta_m$ is the instrument noise with a variance of $\sigma_m^2$. We write a

series of such measurements as an $m_{tot}$ dimensional vector

$$\vec{D} = \vec{\Delta T} + \vec{\Delta}. \tag{1}$$

One can conveniently represent $\vec{\Delta T}$ in matrix form as

$$\begin{pmatrix} \Delta T_1 \\ \Delta T_2 \\ \Delta T_3 \\ \vdots \end{pmatrix} = \overset{\longleftarrow 6144 \longrightarrow}{\begin{pmatrix} 0 & 1 & 0 & 0 & \ldots & 0 & -1 & 0 & 0 & \ldots & 0 \\ 1 & 0 & 0 & 0 & \ldots & 0 & 0 & -1 & 0 & \ldots & 0 \\ 0 & 0 & -1 & 0 & \ldots & 0 & 1 & 0 & 0 & \ldots & 0 \\ \vdots & \vdots & \vdots & \vdots & \vdots & \vdots & \vdots & \vdots & \vdots & \vdots & \vdots \end{pmatrix}} \begin{pmatrix} T_1 \\ T_2 \\ T_3 \\ \vdots \\ T_{6144} \end{pmatrix} \tag{2}$$

or

$$\vec{\Delta T} = \mathbf{V}\vec{T}, \tag{3}$$

where $\mathbf{V}$ is an $m_{tot} \times 6144$ design matrix containing the pointing information and $\vec{T}$ is the sky map we want to solve for. We form a map by minimizing $\chi^2$ defined as

$$\chi^2 = \sum_{m=1}^{m_{tot}} \frac{(D_m - \Delta T_m)^2}{\sigma_m^2}. \tag{4}$$

Setting $\frac{\partial \chi^2}{\partial T_k} = 0$ yields the normal equations

$$\mathbf{A}\vec{T}_{obs} = \vec{M}, \tag{5}$$

where $\vec{T}_{obs}$ is the least-squares estimate of the true temperatures $\vec{T}$, $\mathbf{A} = \mathbf{V}^T \boldsymbol{\Sigma} \mathbf{V}$ where $\boldsymbol{\Sigma}$ is an $m_{tot} \times m_{tot}$ diagonal matrix with $\Sigma_{mm} = 1/\sigma_m^2$, and $\vec{M} = \mathbf{V}^T \boldsymbol{\Sigma} \vec{D}$ is the measurement vector. The DMR instruments have proven to be very stable in flight. Thus we may write $\sigma_m^2 = \sigma_{ch}^2$, where $\sigma_{ch}^2$ is the channel-dependent instrument noise variance. Thus the multiplicative factor $1/\sigma_{ch}^2$ cancels in equation (5) permitting the convenient redefinition of $\mathbf{A}$ to the dimensionless $\mathbf{A} = \mathbf{V}^T \mathbf{V}$.

The calibration, baseline removal and systematic error corrections are performed on $\vec{D}$ before the $\chi^2$ minimization (Janssen & Gulkis 1992). If there are any errors in those procedures that correlate with position on the sky, they will create spurious signals in the maps. A lock-in amplifier memory effect couples $\Delta_m$ with $\Delta_{m-1}$ at the 3.2 % level (Kogut et al. 1992). It is corrected leaving a typical residual coupling of 0.1%.

The matrix $\mathbf{A}$ is $6144 \times 6144$, symmetric, sparse, positive semi-definite and formally singular due to the differential nature of the observations. We solve the normal equations by augmenting the diagonal terms of **A** by a small positive number $\epsilon$. This does nothing more than impose an arbitrary mean level on an otherwise unique solution (Lineweaver 1994). A Gauss-Seidel procedure is used to find an iterative solution to the normal equations and gives the desired best-fit sky map $\vec{T}_{obs}$. For further details on how the maps are made see Torres et al. (1988), Smoot et al. (1990), Keegstra et al. (1991), Jackson et al. (1991), Janssen & Gulkis (1992), and Wright (1994b).

We combine equations (1) and (5) with the definition $\vec{\delta} \equiv \mathbf{V}^T \vec{\Delta}$ to write the best-fit temperatures as a function of the true temperatures plus a noise term $\vec{n}$,

$$\vec{T}_{obs} = \vec{T} + \vec{n} = \vec{T} + \mathbf{A}^{-1} \vec{\delta}. \tag{6}$$

## 3. Correlated Noise Estimates

### 3.1. Correlated Noise Estimates from the Covariance Matrix

The covariance matrix of the sky map temperatures fully describes the noise properties of the map. The diagonal elements are the variances of the fitted temperatures while the non-zero off-diagonal elements represent correlated noise in the maps. We use equation (6) and the uncorrelated nature of the measurement errors to obtain the covariance matrix of the sky map temperatures,

$$Cov(\vec{T}_{obs}) = <\vec{n}\,\vec{n}^T> = [\mathbf{A}^{-1}\mathbf{V}^T]Cov(\vec{\Delta})[\mathbf{A}^{-1}\mathbf{V}^T]^T = \sigma_{ch}^2 \mathbf{A}^{-1}. \tag{7}$$

Thus, the diagonal elements of $\sigma_{ch}^2 \mathbf{A}^{-1}$ are the variances of the fitted temperatures. For 350 randomly chosen values of $i$ we find that $\sigma_{ch}^2 A_{ii}^{-1} = (\sigma_{ch}^2/N_i)(1.007 \pm 0.003)$, where $N_i$ is the number of times pixel $i$ was observed. Thus, the solution of the normal equations produces a 0.7% increase over the variance expected from a single beam experiment $\sigma_{ch}^2/N_i$.

The matrix $\mathbf{A}^{-1}$ is symmetric and non-sparse. Since a column of $\mathbf{A}^{-1}$ has 6144 elements, it can be conveniently viewed as a pixelized map of the correlations of each pixel with the central pixel (i.e. with the diagonal element). The pattern of the correlations is approximately radially symmetric around the central pixel. For each of the 350 randomly chosen columns, we calculate normalized averages of annular regions of the off-diagonal elements

$$A_i^{-1}(\alpha) = \frac{1}{W} \frac{\sum_k^{\alpha_{ik}=\alpha} A_{ik}^{-1}}{A_{ii}^{-1}}, \tag{8}$$

where all pixels in the annuli have constant angular separation from the central pixel and $W = \sum_k^{\alpha_{ik}=\alpha} 1$. The 2.6° pixel size is used as a binning width. Figure 1 shows the average $<A^{-1}(\alpha)>$ and 68% confidence levels of the 350 randomly chosen columns. The mean covariance at 60° is $0.45\%^{+0.18}_{-0.14}$ of the mean variance. Thus the noise correlations in the DMR maps are small and the noise properties resemble those of a single beam experiment. The value of the zero lag bin is off scale; $<A^{-1}(0)> = 1.007$, and is the source of the 0.7% excess variance reported above. This analysis is based on the normal equations from the first year data of channel 53A. Since the sky coverage and the 60° horn separation are only weakly channel dependent, the covariance matrices of the other channels are essentially the same and time independent.

### 3.2. Correlated Noise Estimates from Monte Carlo Simulations

The two-point auto-correlation function of a pixelized DMR map is the weighted average product of temperatures separated by angle $\alpha$

$$C(\alpha) = \frac{1}{W} \sum_{i,j} w_i w_j T_{obs,i} T_{obs,j}, \qquad (9)$$

where the sum is over all pixel pairs $ij$ whose separation angle $\alpha_{ij}$ lies within half a bin width of $\alpha$. We use the weights $w_i = 1/var(T_{obs,i})$ and approximate $var(T_{obs,i})$ by $\sigma_{ch}^2/N_i$. The normalization factor is $W = \sum_{i,j} w_i w_j$. For cross-correlation functions, $i$ refers to one map and $j$ to the other. Using equation (6) the auto-correlation function is defined as

$$C(\alpha) = \frac{1}{W} \sum_{ij}^{\alpha_{ij}=\alpha} w_i w_j (T_i + n_i)(T_j + n_j). \qquad (10)$$

The signal-noise cross-terms in equation (10) contribute to the uncertainty in $C(\alpha)$ but do not contain a bias. Auto-correlation functions of DMR maps are biased only by the noise-noise cross-term. To examine these biases we insert only instrument noise pair-wise into the measurement vector $\vec{M}$ of the normal equations. We then compute the ensemble average of the correlation functions of the resulting maps and obtain,

$$<C(\alpha)> = \frac{1}{W} \sum_{ij}^{\alpha_{ij}=\alpha} w_i w_j <n_i n_j> \qquad (11)$$

$$= \frac{1}{W} \sum_{ij}^{\alpha_{ij}=\alpha} w_i w_j \sigma_{ch}^2 A_{ij}^{-1}. \qquad (12)$$

– 7 –Figure 2a shows the bias in the auto-correlation function from 1000 simulations of single-year 53A noise maps. One can see a positive bias of $\sim 1.5\,\sigma$ at 60°, where $\sigma$ is the half-width of the gray band, and a smaller positive bias close to 0°. The bias does not change substantially when a signal is included in the simulations. Figure 2b shows that cross-correlations (e.g. Figure 3 of Smoot et al. 1992) do not suffer from the pixel-pixel noise correlations.

The positive correlation between noise in pixels separated by 60° can be explained as follows: the total noise in each pixel *includes* contributions from the off-diagonal elements of the covariance matrix $\sum_j A_{ij}^{-1}\delta_j$, $j \neq i$. These contributions are dominated by the average temperature of the noise in the reference ring at 60° angular separation. Thus the central pixel is *positively* correlated with the reference ring.

The size of the biases in Figure 2a scale with the variance of the noise in the maps. For example, the sizes of the biases in the auto-correlations of two year maps are half the sizes of the biases in the one year maps.

### 3.3. Correlated Noise Estimates from DMR Maps

An estimate of the bias can be obtained using only the data. We take advantage of the previous result that the auto-correlations contain the bias while the cross-correlations do not. Their difference is then an estimate of the bias. We introduce the quantity $B(\alpha)$ as an empirical estimate of the bias defined as

$$B(\alpha) = \frac{1}{12} \sum_{m=1}^{6} \left[ \frac{C_{A,m}(\alpha) - C_{A\times B,m}(\alpha)}{\sigma_{A,m}(\alpha)} + \frac{C_{B,m}(\alpha) - C_{A\times B,m}(\alpha)}{\sigma_{B,m}(\alpha)} \right], \qquad (13)$$

where $C_{A,m}, C_{B,m}$ and $C_{A\times B,m}$ are respectively, the auto-correlation of channel A, auto-correlation of channel B and their cross-correlation, for 6 values of $m$ (2 years $\times$ 3 frequencies). The empirical bias $B(\alpha)$ is expressed in dimensionless units of noise-only standard deviations of the auto-correlations and is plotted in Figure 3. The bias at 60° is evident.

### 4. Cosmological Implications

The noise correlations at 60° have the effect of putting additional small angular scale ($\ell \gtrsim 6$) power into the maps. The resultant correlation function more closely resembles a correlation function from a higher power spectral index $n$ and lower amplitude $Q_{rms-PS}$. We obtain biases on $n$ and $Q_{rms_{PS}}$ from $\chi^2$ fits including cosmic variance (Smoot et al. 1992) to the mean of the auto and cross-correlations of 200 simulations of one year 53A correlated noise superimposed on $n = 1$, $Q_{rms-PS} = 17 \mu K$ signal. Correlated noise increases $n$ by $\sim 0.2$ and decreases $Q_{rms-PS}$ by $\sim 1.2$ $\mu K$. When cosmic variance dominates the errors in $n$ and $Q_{rms-PS}$, the correlated noise biases in auto-correlation functions become less important since the biases scale with the variance of the noise. Determinations of $n$ and $Q_{rms-PS}$ from cross-correlation functions (Smoot et al. 1992, Bennett et al. 1994) and cross-power spectra (Wright et al. 1994c) are not affected by these noise correlations.

The bias in Figure 2a near $\alpha = 0°$ increases the amplitude of rms temperature fluctuations determined from the first non-zero bin of auto-correlations. The $C(0)^{1/2} = 36 \pm 5$ $\mu K$ quoted in Bennett et al. (1994) is from the zeroth bin of the 53 × 90 GHz cross-correlation and is not susceptible to noise correlation bias. The correlated noise near $\alpha = 0°$ does not correlate across channels and therefore contributes the same to both the (A+B)/2 and (A-B)/2 maps. Therefore $\sigma_{sky} = \sqrt{\sigma^2_{A+B} - \sigma^2_{A-B}}$ is unbiased and the previously published values of rms temperature fluctuations (Smoot et al. 1992, Bennett et al. 1994) are not affected by correlated noise.

## 5. Summary

We have investigated the structure and magnitude of the correlated noise in DMR maps using the covariance matrix of the pixel temperatures, two-point correlation functions of Monte Carlo simulations and the output maps themselves. The noise in the DMR maps resembles that from a single beam experiment in that the pixel-pixel noise correlations are small. The approximately radial structure of the pixel noise covariance matrix is shown in Figure 1. The dominant correlation in each pixel is from the average noise in a ring 60° away. The mean covariance at 60° is 0.45% of the mean variance. The noise variance is 0.7% larger than the expected variance. For auto-correlation functions at $\alpha = 60°$ there is a $\sim 1.5$ $\sigma$ positive bias. Determinations of $n$ and $Q_{rms-PS}$ from auto-correlations are biased towards higher $n$ and lower $Q_{rms-PS}$ values. Cross-correlation functions are not affected by noise correlations and should be used for $n$, $Q_{rms-PS}$ and $C(0)^{1/2}$ determinations, thus previously published DMR results are not significantly affected by correlated noise.

The COBE project has been supported by the Astrophysics Division of the NASA Office of Space Sciences and Applications. We thank J. Aymon, G. De Amici, P. Jackson, E. Kaita, V. Kumar, R. Kummerer, K. Loewenstein and C. Witebsky for their work in suppport of the DMR.

---

# Captions

Figure 1
The average radial structure of the pixel temperature covariance matrix $\mathbf{A}^{-1}$ (see equation (8)). The shaded band marks the 68% confidence levels.

Figure 2
a) The 2-point auto-correlation of 1000 simulated 53A single-year noise maps. A 20° Galactic cut has been made. The best-fit mean, dipole and quadrupole have been removed from each simulation. Weighted $C(\alpha)$'s were used. The shaded band marks the 68% confidence levels. b) Cross-correlations from the same simulated noise maps as (a). No bias is evident.

Figure 3
The auto-correlation function bias with respect to the corresponding cross-correlations. This empirical bias $B(\alpha)$ is computed from twelve independent DMR data sets (2 years × 3 frequencies × 2 channels). See equation (13) of text for the definition of $B(\alpha)$. The shaded band marks the 95% confidence levels from 250 simulations with uncorrelated noise. The best-fit mean, dipole and quadrupole have been removed before $B(\alpha)$ was calculated. The result is not strongly dependent on which multipole is removed. Weighted $C(\alpha)$'s were used.

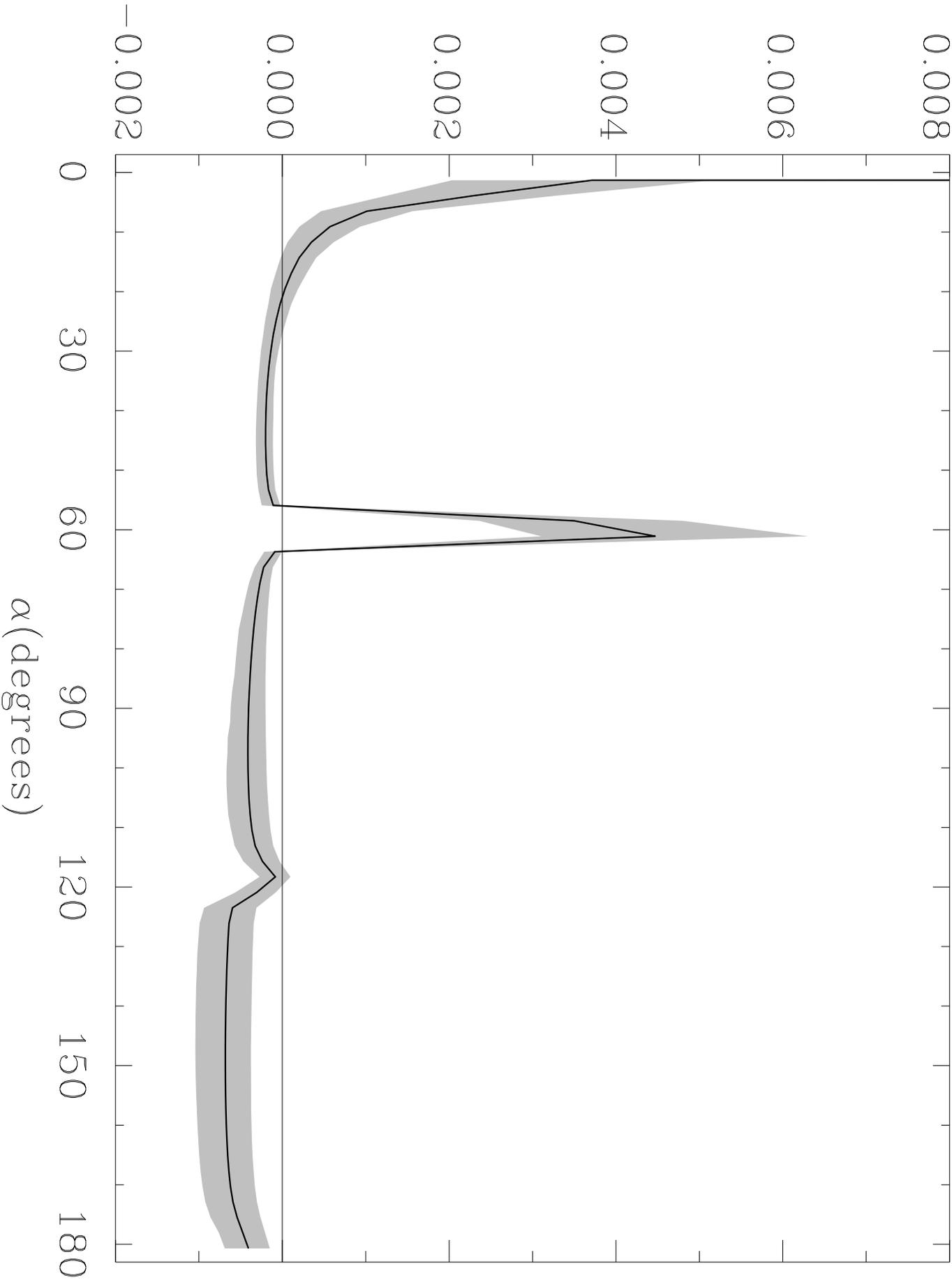

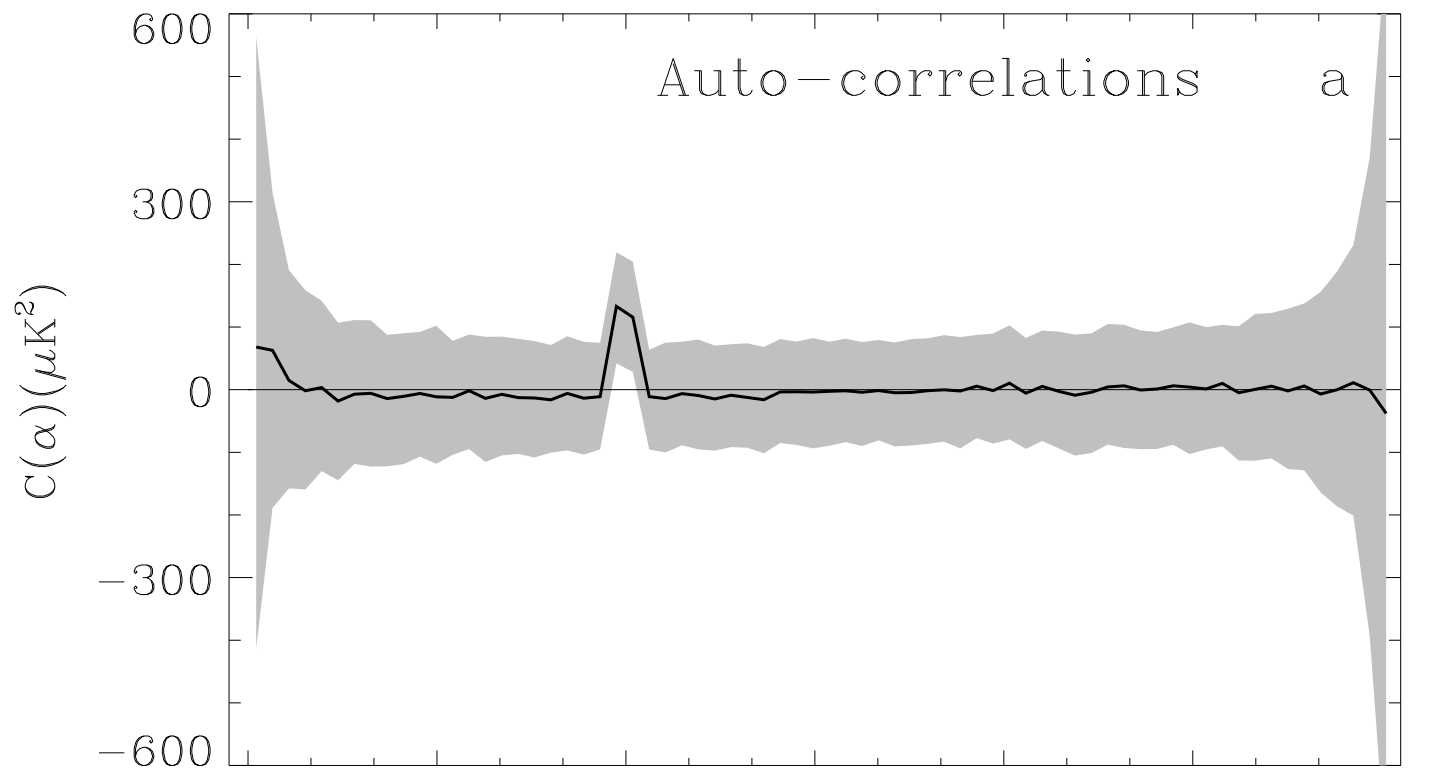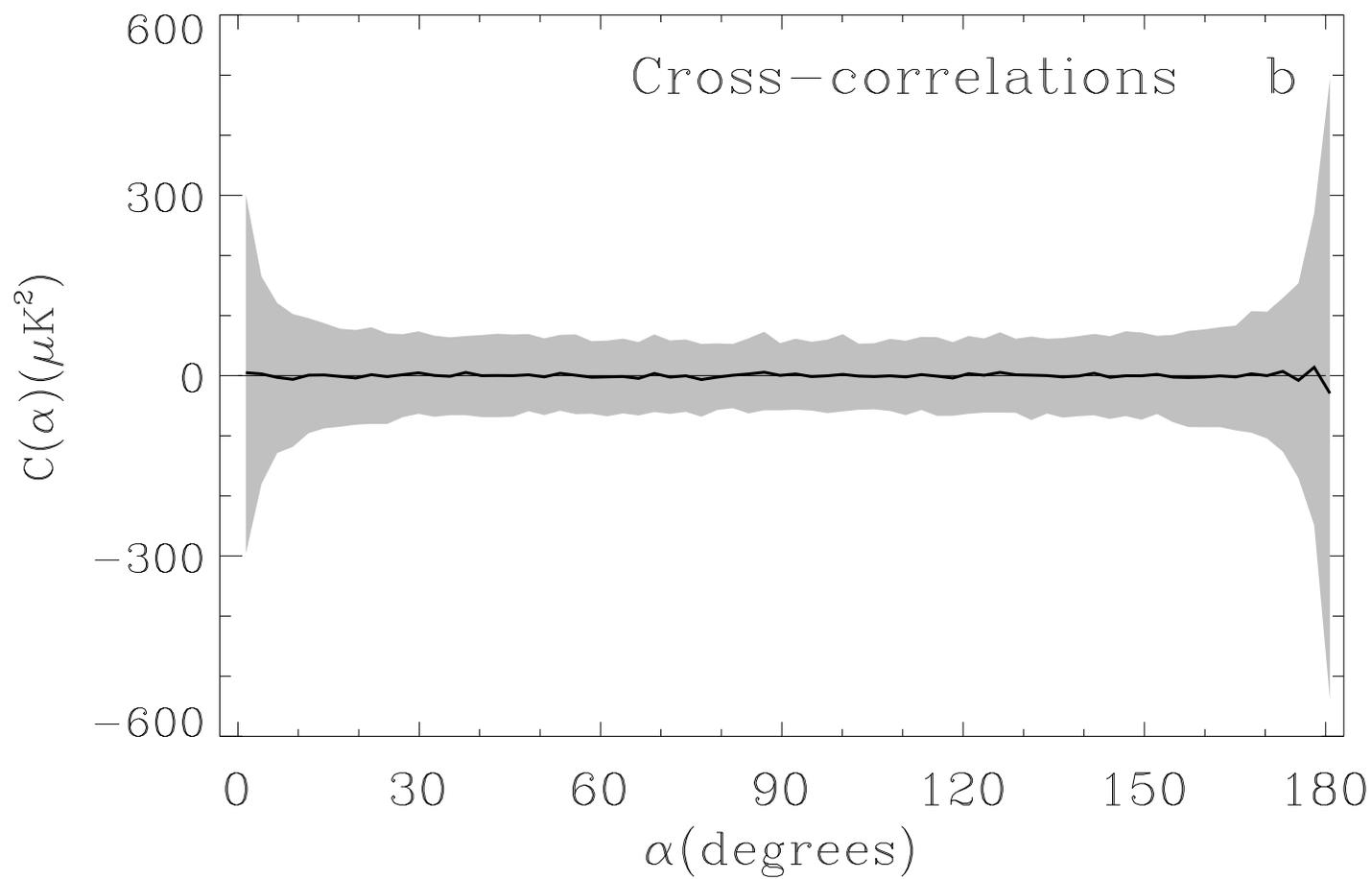

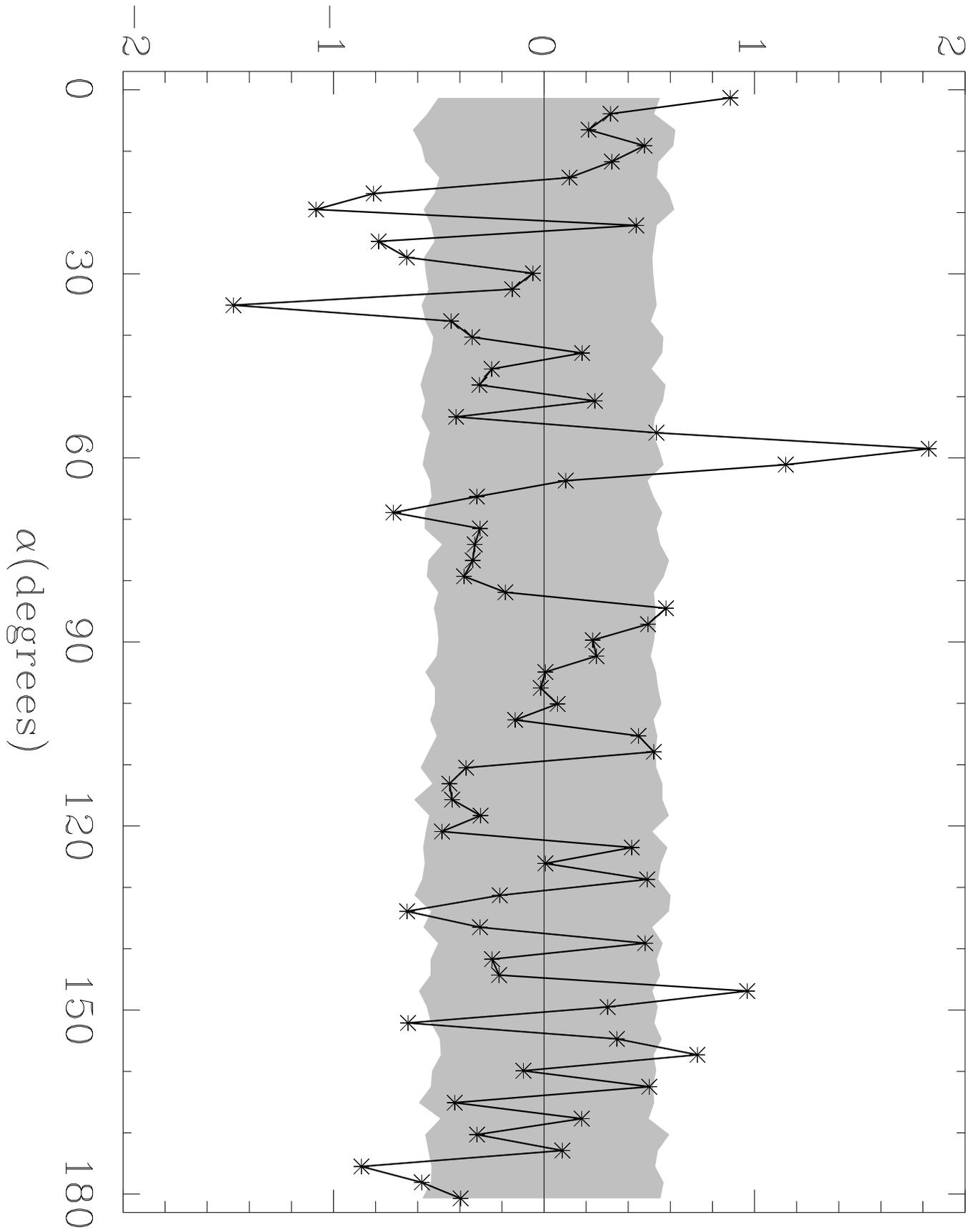